\documentclass{jpsj-suppl}
\usepackage{times}
\usepackage{bm}
\usepackage{ulem}

\begin{document}

\title{Charge Order with a Noncoplanar Triple-$Q$ Magnetic Order on a Cubic Lattice}

\author{Satoru Hayami\thanks{E-mail address: hayami@aion.t.u-tokyo.ac.jp}, Takahiro Misawa, and Yukitoshi Motome 
}
\inst{Department of Applied Physics, University of Tokyo, Hongo 7-3-1, Bunkyo, Tokyo 113-8656 
}
\abst{
Ground state 
of the periodic Anderson model on a cubic lattice 
is investigated by mean-field calculations, with focusing on the possibility of charge ordering.
We show that 
three different charge-ordered phases 
appear at 3/2 filling, which are accompanied by ferromagnetic, 
noncoplanar triple-$Q$, and ferrimagnetic order, respectively. 
The triple-$Q$ magnetic order is likely stabilized by 
emergent geometrical frustration 
in the charge-ordered phase with the ordering wave vector $(\pi,\pi,\pi)$. 
We show that 
the nearest-neighbor hopping between localized electrons 
stabilizes the triple-$Q$ phase relative to the other two phases. 
We also discuss the electronic structure in the triple-$Q$ charge-ordered phase. 
}

\kword{charge order, periodic Anderson model, cubic lattice, 
triple-$Q$ magnetic order, emergent geometrical frustration}

\maketitle

\section{Introduction}
Heavy-fermion systems are 
prototypical systems 
where the interaction 
between two different types of electrons,~i.e., conduction and localized electrons, 
leads to fascinating behavior, 
such as non-Fermi liquid behavior, quantum criticality, and unconventional superconductivity~\cite{RevModPhys.56.755,mathur1998magnetically,RevModPhys.73.797}. 
To understand the nature of heavy-fermion systems, 
the periodic Anderson model and the Kondo lattice model have been extensively studied 
as the fundamental models to describe 
interplay between conduction and localized electrons~\cite{Hewson199704,PhysRev.124.41,RevModPhys.69.809}. 

Although the periodic Anderson model and the Kondo lattice model 
include neither bare off-site repulsive Coulomb interactions 
nor electron-phonon interactions, a possibility of charge order (CO) in these models 
has been discussed intensively. 
In contrast to the conventional CO, which is usually driven by
the repulsive Coulomb interactions and/or 
the Peierls instability~\cite{PhysRevB.17.494,RevModPhys.60.1129}, 
Hirsch first pointed out that effective repulsive interactions 
are induced in the limit of strong Kondo coupling, which might 
lead to a CO state~\cite{PhysRevB.30.5383}. 
However, a signature of such CO was not found 
in the one-dimensional Kondo lattice model~\cite{RevModPhys.69.809, PhysRevB.65.052410,PhysRevB.84.115116}. 
Recently, in infinite dimensions, 
it was claimed that CO appears in the Kondo lattice model 
by using the dynamical mean-field theory~\cite{JPSJ.78.034719,PhysRevB.87.165133}.
In addition, two of the authors and their collaborator 
showed that CO with N\'eel-type antiferromagnetic order 
is stabilized in the two-dimensional Kondo lattice model on a square lattice 
by using the cellular dynamical mean-field theory and 
the variational Monte Carlo method~\cite{PhysRevLett.110.246401}. 
Meanwhile, in the periodic Anderson model, 
the possibility of ferromagnetic CO was 
discussed  in infinite and three dimensions~\cite{Zadeh_PhysRevB.55.R3332, Majidi_2007charge}. 
CO was also studied on geometrical frustrated lattices; 
for example, on a triangular lattice, CO was discussed associated with 
partial magnetic disorder in the Kondo lattice model~\cite{Motome2010,PhysRevLett.108.257205} and the periodic Anderson model~\cite{Hayami2011,Hayami2012}.

Although CO in one, two, and infinite dimensions were studied in detail,
the possibility in realistic three-dimensional systems has not been studied systematically except for the ferromagnetic CO state.
Specifically, it is interesting to ask whether a CO state with the N\'eel-type antiferromagnetic order is also realized in three dimensions. 
As we will show later, it turns out that 
magnetic frustration emerges under three-dimensional CO, which makes the magnetism in the CO phases complicated.
In order to clarify the possibility of CO, 
it is necessary to perform the systematic calculations which treat various magnetic states on an equal footing.

In the present study, we explore CO states 
on a three-dimensional cubic lattice. 
Especially, we focus on competition and cooperation between CO and magnetic orders. 
By investigating the ground state 
of the periodic Anderson model at 3/2 filling by 
the mean-field approximation, 
we find that the model exhibits three different CO states. 
Despite the absence of apparent frustration in the cubic lattice, 
one of the CO states shows a noncoplanar triple-$Q$ magnetic order. 
We show that the origin 
is likely to be the effective geometrical frustration induced by CO with the ordering wave vector $(\pi,\pi,\pi)$ 
on the cubic lattice. 
We also examine the electronic structure for the CO state with the noncoplanar magnetic order. 

\section{Model and Method}

We consider the periodic Anderson model, whose 
Hamiltonian is given by 
\begin{eqnarray}
\label{PAM_Ham}
{\mathcal{H}}   &=& -  t \sum_{\langle i,j\rangle,\sigma} 
( c^{\dagger}_{i \sigma} c_{j \sigma}  + {\mathrm{H.c.}} ) 
-  t_f \sum_{\langle i,j\rangle,\sigma} 
( f^{\dagger}_{i \sigma} f_{j \sigma}  + {\mathrm{H.c.}} ) 
- V\sum_{i ,\sigma} 
( c^{\dagger}_{i \sigma}f_{i \sigma}+{\mathrm{H.c.}} ) \nonumber  \\
 & &+U \sum_{i} f^{\dagger}_{i \uparrow}f_{i \uparrow} f^{\dagger}_{i \downarrow} f_{i \downarrow} 
+ E_{0} \sum_{i, \sigma} f^{\dagger} _{i \sigma} f_{i \sigma}. 
\end{eqnarray} 
Here, $c_{i\sigma}^{\dagger} (c_{i \sigma}) $ and $f_{i\sigma}^{\dagger} (f_{i \sigma}) $ are 
the creation (annihilation) operators of conduction and localized electrons 
at site $i$ and spin $\sigma$, respectively. 
The first (second) term in Eq.~(\ref{PAM_Ham}) represents 
the kinetic energy of conduction (localized) electrons 
(we assume that $t>0$, $t_f \geq 0$, and $t_f$ is considerably smaller than $t$), 
the third term the onsite $c$-$f$ hybridization 
between conduction and localized electrons ($V>0$), 
the fourth term the onsite Coulomb interaction for localized electrons ($U>0$), 
and the last term the atomic energy of localized electrons. 
The sum of $\langle i, j \rangle$ is taken over the nearest-neighbor sites on the cubic lattice. 
Hereafter, we take $t=1$ as an energy unit, 
the lattice constant $a=1$, and $E_0=-4$. 
We focus on a commensurate filling, 
$n =(1/N)\sum_{i \sigma} \langle c_{i \sigma}^{\dagger}c_{i \sigma} 
+ f_{i \sigma}^{\dagger} f_{i \sigma}  \rangle = 3/2$, 
where $N$ is the total number of sites. 
We note that this filling corresponds to 1/4 filling 
in the Kondo lattice model in the large $U$ limit 
where the $f$ level is singly occupied to form a localized moment at each site. 
In the Kondo lattice model, 
CO was found in the intermediate Kondo coupling region at and around 1/4 filling~\cite{JPSJ.78.034719,PhysRevB.87.165133,PhysRevLett.110.246401}. 

In order to obtain the ground state of the model in Eq.~(\ref{PAM_Ham}), 
we apply the standard Hartree-Fock 
approximation, which preserves SU(2) symmetry of the model, to the Coulomb $U$ term. 
In the calculations, we employ 
the $2\times 2\times 2$-site unit cell. 
We also confirm that 
the dominant phases are 
qualitatively unchanged in the calculations 
with using the $4\times 4\times 4$-site unit cell. 
We calculate the mean fields 
$\{ \langle f_{i \sigma}^{\dagger} f_{i \sigma' } \rangle \}$ by taking the sum over 
$16\times 16\times 16$ grid points in the first Brillouin zone. 
The mean fields are determined self-consistently, 
by repeating calculations until 
each mean field converges within the precision less than $10^{-5}$. 

\section{Result and Discussion}

\begin{figure}
\begin{center}
\includegraphics[width=1.0 \hsize]{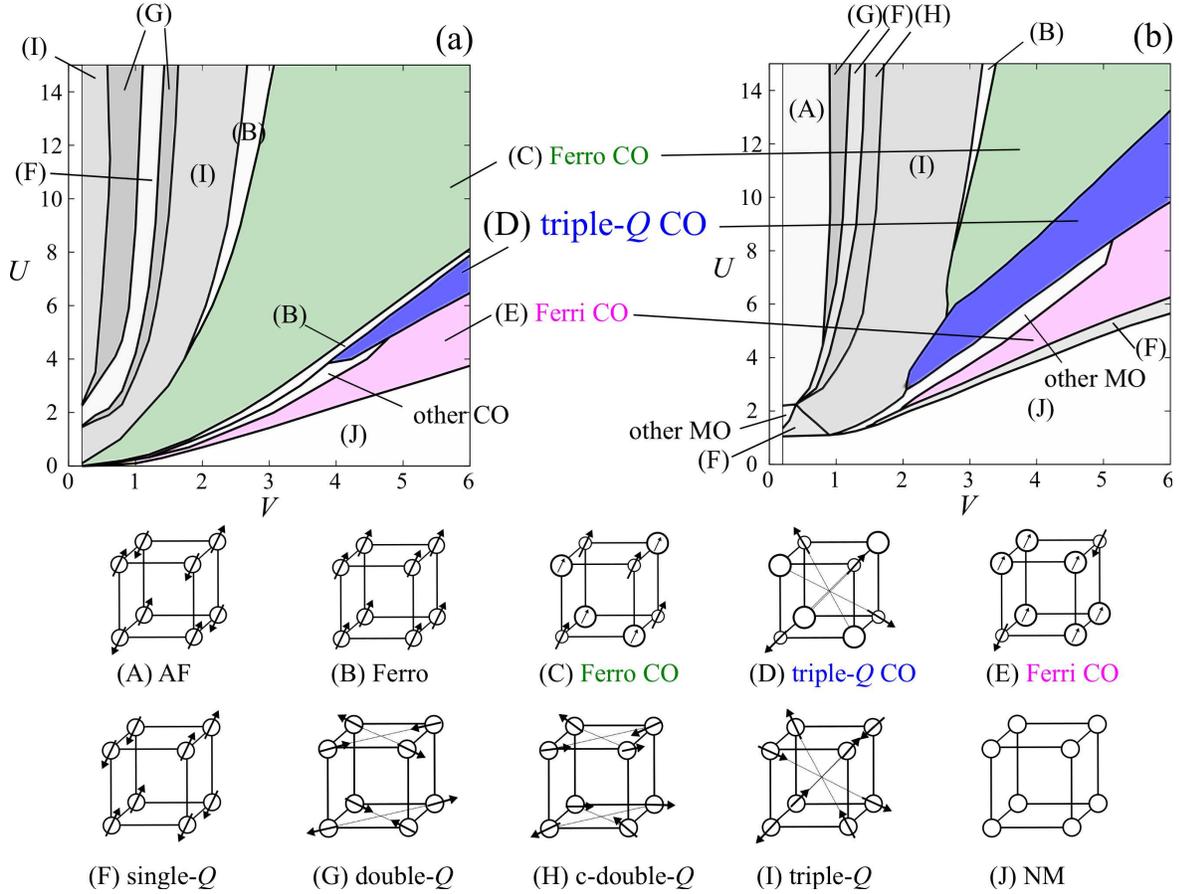}
\end{center}
\caption{(Color online). 
Ground-state phase diagram of the periodic Anderson model [Eq.~(\ref{PAM_Ham})] 
at 3/2 filling obtained by the mean-field calculations for (a) $t_f=0.0$ 
and (b) $t_f=0.2$. 
(b) is the same as Fig.~2 in Ref.~\citen{hayami20133d}. 
Schematic pictures of the ordering patterns in $f$ electrons are shown in the bottom panel. 
The sizes of circles denote local electron densities, and the arrows represent local spin moments. 
AF, Ferro, Ferri, CO, and NM stand for 
antiferromagnetic (N\'eel-type), ferromagnetic, ferrimagnetic, 
charge-ordered, and nonmagnetic metallic states, respectively. 
Single-$Q$ corresponds to $\bm{Q} = (\pi, 0, \pi)$, 
double-$Q$ $(0, \pi, \pi)$, $(\pi, 0, \pi)$, and
triple-$Q$ $(\pi, \pi, 0)$, $(0, \pi, \pi)$, $(\pi, 0, \pi)$. 
C-double-$Q$, other CO, and other MO represent the canted double-$Q$, 
other charge-ordered states, and other magnetically ordered states, respectively. 
}
\label{fig1}
\end{figure}

Figure~\ref{fig1}(a) shows the ground-state phase diagram 
obtained by the mean-field calculations for $t_f = 0$. 
The phase diagram includes various phases, which is roughly divided into three regions: 
the large-$U$ small-$V$ region where magnetic phases with multiple-$Q$ ordering are dominant, 
the intermediate $U$ and $V$ 
region where CO phases are dominant, and 
the small-$U$ large-$V$ region where the nonmagnetic phase appears. 
Schematic pictures of magnetic and CO 
states for $f$ electrons are shown in the bottom panel of Fig.~\ref{fig1}. 

Among the various phases, 
here, we focus on three CO phases in the intermediate $U$ and $V$ region: 
the ferromagnetic, triple-$Q$, and ferrimagnetic CO phases. 
We first discuss the result for the ferromagnetic CO, which is robustly 
stabilized in the wide region for intermediate $U$ and $V$. 
This state is insulating and CO is characterized 
by the ordering wave vector $(\pi, \pi, \pi)$. 
The magnitude of magnetic moments for localized electrons, 
$m_{i}^{f} = [\langle {s}_{i,x}^{f} \rangle^2 +\langle {s}_{i,y}^{f} \rangle^2 
+ \langle {s}_{i,z}^{f} \rangle^2]^{1/2}$, at charge-rich sites 
is typically 0.8-0.9 times smaller than that at charge-poor sites 
(${s}_{i,\mu}^f$ is the $\mu$ component of the spin operator for $f$ electron). 
Similar ferromagnetic CO states were found also in the previous 
studies in infinite dimensions~\cite{Zadeh_PhysRevB.55.R3332} 
and three dimensions~\cite{Majidi_2007charge}. 
For larger $V$, the charge disproportionation becomes smaller and 
continuously vanishes at the phase boundary to 
the ferromagnetic phase without CO, which is denoted by the phase (B) in Fig.~\ref{fig1}(a). 

While further increasing $V$, 
the first-order phase transition from the ferromagnetic state to 
the triple-$Q$ CO state occurs around 
$V \gtrsim 4$ and $U \gtrsim 4$, as shown in Fig.~\ref{fig1}(a). 
The triple-$Q$ spin configuration on the cubic lattice is noncoplanar; namely, it is represented 
by $\langle \bm{s}_{i}^{f} \rangle =(\langle {s}_{i,x}^{f} \rangle, \langle {s}_{i,y}^{f} \rangle, \langle {s}_{i,z}^{f} \rangle) 
= m [\cos(\bm{Q}_1\cdot \bm{r}_{i}),\cos(\bm{Q}_2\cdot \bm{r}_{i}),\cos(\bm{Q}_3 \cdot \bm{r}_{i})]$ (see Fig.~\ref{fig1}). 
Here, $m$ is the magnetic order parameter, ${\bm r}_{i}$ is the position vector of the site $i$, and 
${\bm Q}_{1} = (\pi, 0, \pi)$, ${\bm Q}_{2} = (0, \pi, \pi)$, and ${\bm Q}_{3} = (\pi, \pi, 0)$. 
The CO pattern is similar to that in the ferromagnetic CO state. 
The magnetic moment at the charge-rich site ($m_{\rm rich}^{f}$) is 
negligibly small compared to that at the charge-poor site ($m_{\rm poor}^{f}$); for instance, 
$m_{\rm poor}^{f} \sim 0.2$ 
and $m_{\rm rich}^{f} < 0.01$ at $U=6$ and $V=5$. 
This triple-$Q$ CO state is an insulator for large $U$, 
while it turns into a metal with decreasing $U$, as discussed later. 
For a larger $V$, the first-order phase transition occurs 
from the triple-$Q$ CO state to the ferrimagnetic CO state. 
The CO pattern is different from that in the ferromagnetic and triple-$Q$ CO states. 
As shown in Fig.~\ref{fig1}, it is the four-sublattice order, in which 
three of four sites are charge rich. 
The magnetic structure is an up-up-up-down collinear type and the spin structure factor has four peaks 
at $(\pi, \pi, 0)$, $(0, \pi, \pi)$, $(\pi, 0, \pi)$, and $(0, 0, 0)$. 
The ferrimagnetic CO state is a metal.

\begin{figure}
\begin{center}
\includegraphics[width=1.0 \hsize]{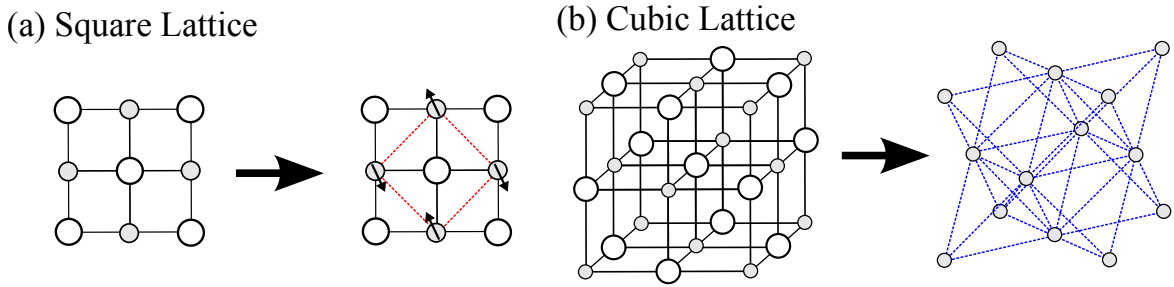}
\end{center}
\caption{(Color online). 
Schematic pictures of the CO states on (a) the square lattice and (b) the cubic lattice. 
The small (large) circles represent charge-poor(rich) sites, and the arrows 
on the charge-poor sites denote the directions of local spins. 
The dashed lines connect 
the nearest-neighbor charge-poor sites (the next-nearest-neighbor sites on the original lattices). 
In (a), the charge-poor sites form the 45$^{\circ}$-rotated square lattice with 
$\sqrt{2}\times\sqrt{2}$-times larger unit cell, 
while in (b), they consist of the face-centered-cubic lattice. 
}
\label{fig2}
\end{figure}

Among the three CO phases, the triple-$Q$ CO phase is most interesting. 
In fact, the CO 
phase with noncoplanar spin state was not seen in the two-dimensional square lattice~\cite{PhysRevLett.110.246401}. 
Here, we explain why the triple-$Q$ CO state appears in the cubic lattice case. 
The difference of magnetism in the CO states between the square- and cubic-lattice systems 
is understood by considering the geometry of the dominantly-magnetic charge-poor sites. 
In the case of the square lattice, 
the charge-poor sites comprise a 45$^{\circ}$-rotated square lattice with $\sqrt{2} \times \sqrt{2}$-times larger unit cell, 
as schematically shown in Fig.~\ref{fig2}(a); 
there is no frustration in the antiferromagnetic interaction 
between the nearest-neighbor charge-poor sites, and hence, 
the collinear antiferromagnetic order develops along with CO. 
On the other hand, in the case of the cubic lattice, 
the charge-poor sites form the face-centered-cubic lattice, as shown in Fig.~\ref{fig2}(b). 
Thus, CO on the cubic lattice induces the effective geometrical frustration 
between the magnetic moments at charge-poor sites. 
This may suppress a development of simple antiferromagnetic orders, and opens a possibility of 
noncollienar or noncoplanar magnetic ordering. 
We note that the triple-$Q$ state is energetically degenerate with other spin configurations, such as 
an up-up-down-down collinear state, in the classical Heisenberg spin system with the nearest-neighbor 
antiferromagnetic interaction on the face-centered-cubic lattice. 
In the spin-only system, it was pointed out that the degeneracy is lifted by the higher-order exchange interactions and the triple-$Q$ state may be selected as the ground state~\cite{0305-4608-13-10-006}. 
In the periodic Anderson model studied here, 
such higher-order contributions can be induced by the kinetic motion of electrons and may play a role in 
stabilizing the noncoplanar triple-$Q$ state. 

Now, we discuss the effect of the nearest-neighbor $f$-electron hopping $t_f$. 
We find that $t_f$ stabilizes the triple-$Q$ CO state relative to other CO states. 
As shown in the ground-state phase diagram at $t_f=0.2$ in Fig.~\ref{fig1}(b), 
the region of the triple-$Q$ CO phase becomes wider, whereas 
the ferromagnetic CO region is reduced and 
the ferrimagnetic one is almost unchanged. 
This tendency is understood by considering the
effective interaction induced by $t_f$ between the nearest-neighbor sites. 
Since in this large $U$ region $f$ electrons are almost at half filling and localized at each site, 
the effective magnetic interactions induced by the second-order perturbation in terms of $t_f$ become antiferromagnetic.
This antiferromagnetic interaction destabilizes the ferro CO phase relative to the triple-$Q$ phase.  
Thus, $t_f$ stabilizes the triple-$Q$ CO state compared to the ferromagnetic one.

\begin{figure}
\begin{center}
\includegraphics[width=0.8 \hsize]{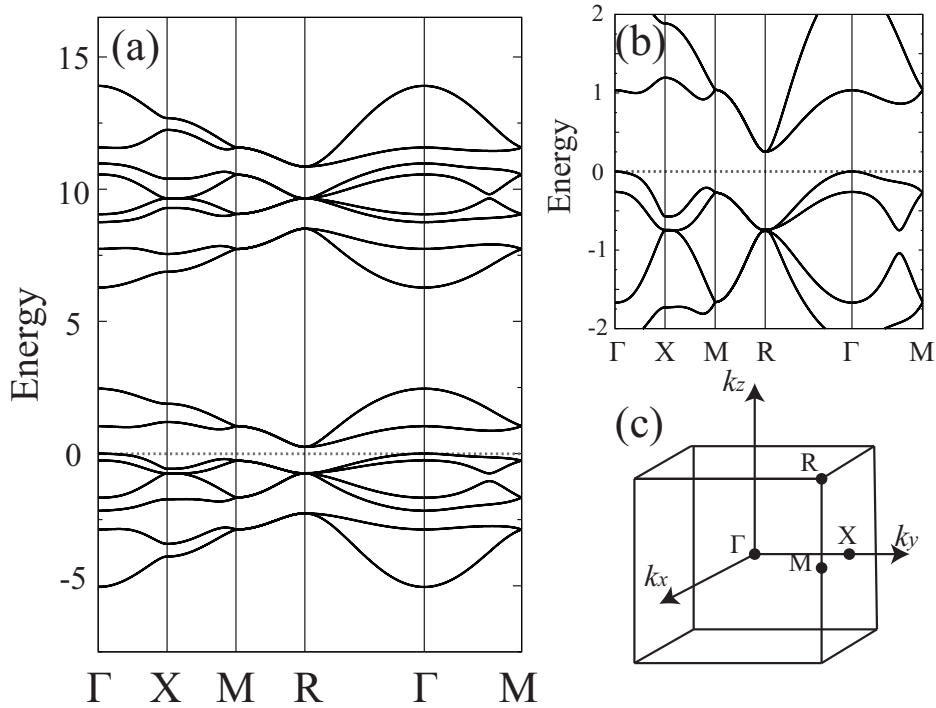}
\end{center}
\caption{ 
Energy dispersion in the triple-$Q$ CO state at 3/2 filling. 
The 
results are calculated at $U=10$, $V=5.2$, and $t_f=0.2$: 
(a) overall structure of the energy dispersion and (b) the energy dispersion in the vicinity of the Fermi level. 
The dashed lines show the energy of the highest occupied level. 
The energy dispersions are shown along the symmetric lines in the magnetic Brillouin zone represented in (c), by connecting the $\Gamma$ $(0, 0, 0)$, X $(0,\pi/2,0) $, M $(\pi/2, \pi/2,0)$, and R $(\pi/2,\pi/2,\pi/2)$ points. 
}
\label{fig3}
\end{figure}

Let us discuss the electronic structure in the triple-$Q$ CO state. 
Figure~\ref{fig3}(a) shows the typical band structure in the triple-$Q$ CO phase. 
The band structure is drawn along the symmetric lines in 
the folded Brillouin zone, i.e., $\Gamma$-X-M-R [see Fig.~\ref{fig3}(c)]. 
Because of the eight-sublattice order, there are totally sixteen bands (doubly degenerate each). 
The energy is measured from the highest occupied level. 
As shown in Fig.~\ref{fig3}(a), the overall band structure 
consists of the bonding and antibonding states, 
which are separated by $\sim 2V$. 
In Fig.~\ref{fig3}(b), we show the band dispersions near the Fermi level. 
The triple-$Q$ CO state has a finite energy gap for this parameter setting ($U=10$, $V=5.2$, and $t_f=0.2$), 
which is the indirect one between the $\Gamma$ and R points. 
As described above, however, there is 
a metal-insulator transition in the triple-$Q$ CO phase while changing $U$. 
As decreasing $U$, the band dispersions around the R point lower and 
eventually, the energy gap closes. 
Hence, while decreasing $U$, the triple-$Q$ CO state changes from insulator to metal, 
which has an electron pocket around the R point 
and a hole pocket around the $\Gamma$ point.

Related to the noncoplanar magnetic ordering in the triple-$Q$ CO state, 
it is interesting to consider the possibility of anomalous Hall effect. 
The triple-$Q$ order that we found here is the same as that in the previous study 
for the double-exchange model on the face-centered-cubic lattice~\cite{Shindou_PhysRevLett.87.116801}. 
It was shown that the quantum anomalous Hall effect 
is induced by the triple-$Q$ order under the uniaxial $\langle 111 \rangle$ lattice distortion. 
Therefore, a similar quantum anomalous Hall effect 
will appear in the current triple-$Q$ CO state by introducing the lattice distortion. 

\section{Summary}
To summarize, we have explored charge-ordered states 
in the periodic Anderson model on a three-dimensional cubic lattice by the Hartree-Fock approximation. 
We have elucidated the ground-state phase diagram at 3/2 filling and
found that the charge order is stabilized with three magnetic orders: 
ferromagnetic, triple-$Q$ magnetic, and ferrimagnetic orders. 
Among them, the triple-$Q$ charge-ordered phase is a 
characteristic state, in which the effective magnetic frustration 
induced by the underlying charge order is likely to play an important role in stabilizing the noncoplanar magnetic order. 
In addition, we have shown that the hopping between localized electrons 
stabilizes the triple-$Q$ charge-ordered phase relative to the other charge-ordered phases.
We have discussed the electronic structure in the triple-$Q$ 
charge-ordered state and the insulator-metal transition.
We have also discussed the possibility of 
quantum anomalous Hall effect in the triple-$Q$ charge-ordered state. 

The authors thank Junki Yoshitake for 
helpful comments on the 
charge-ordered state in three dimensions. 
S.H. is supported by Grant-in-Aid for JSPS Fellows. 
This work was supported by Grants-in-Aid for Scientific Research (Nos. 22540372, 23102708, and 24340076), 
the Strategic Programs for Innovative Research (SPIRE), MEXT, 
and the Computational Materials Science Initiative (CMSI), Japan. 

\bibliographystyle{JPSJ}
\bibliography{ref}

\end{document}